\documentclass[twocolumn,floatfix]{revtex4}
\usepackage{epsf,graphicx}
\usepackage{amsmath,amssymb,amsfonts}
\newcommand{\vecbm}[1]{\mbox{\boldmath#1}}
\newcommand{\vecb}[1]{\mbox{\bf#1}}
\newcommand{\cent}[1] {\begin{center}#1\end{center}}

\newcommand{\nvec}[1]{\stackrel{\rightarrow}{#1}}
\newcommand{\goo}{\,\raisebox{-.5ex}{$\stackrel{>}{\scriptstyle\sim}$}\,}
\newcommand{\stdef} {\stackrel{\mbox{\scriptsize def}}{=}}

\begin{document}
\title{On the foundation of thermodynamics by microcanonical thermostatistics.\\
 The microscopic origin of condensation and phase separations.}
\author{D.H.E. Gross}
\affiliation{Hahn-Meitner Institute and Freie Universit{\"a}t Berlin,
Fachbereich Physik. Glienickerstr. 100; 14109 Berlin, Germany} \email{
gross@hmi.de} \homepage{http://www.hmi.de/people/gross/}

\begin{abstract}
Conventional thermo-statistics address infinite homogeneous systems within
the canonical ensemble. However, some 170 years ago the original motivation
of thermodynamics was the description of steam engines, i.e. boiling water.
Its essential physics is the separation of the gas phase from the liquid.
Of course, boiling water is inhomogeneous and as such cannot be treated by
conventional thermo-statistics. Then it is not astonishing, that a phase
transition of first order is signaled canonically by a Yang-Lee
singularity. Thus it is only treated correctly by microcanonical
Boltzmann-Planck statistics. This is elaborated in the present paper. It
turns out that the Boltzmann-Planck statistics is much richer and gives
fundamental insight into statistical mechanics and especially into entropy.
This can be done to a far extend rigorously and analytically. The deep and
essential difference between ``extensive'' and ``intensive'' control
parameters, i.e. microcanonical and canonical statistics, is exemplified by
rotating, self-gravitating systems. In this paper the necessary appearance
of a convex entropy $S(E)$ and the negative heat capacity at phase
separation in {\em small as well macroscopic systems independently of the
range of the force} is pointed out. The appearance of a critical end-point
for the liquid-gas transition in the $p-E$ or $V-E$ phase diagram can be
easily explained as well the non-existence of a critical end-point of the
solid-liquid transition.
\end{abstract}

\maketitle

\section{Introduction}
Since the beginning of thermodynamics in the first half of the
19.century its original motivation was the description of steam engines and
the liquid to gas transition of water. Here water becomes inhomogeneous and
develops a separation of the gas phase from the liquid, i.e. water boils.

A little later statistical mechanics was developed by
Boltzmann\cite{boltzmann1872} to explain the microscopic mechanical basis
of thermodynamics. Up to now it is generally believed that this is given by
the Boltzmann-Gibbs canonical statistics. As traditional canonical
statistics works only for homogeneous, infinite systems, phase separations
remain outside of standard Boltzmann-Gibbs thermo-statistics, which,
consequently, signal phase-transitions of first order by Yang-Lee
singularities.

It is amusing that this fact that is essential for the original purpose of
thermodynamics to describe steam engines was never treated completely in
the past 150 years.  The system must be somewhat artificially split into
(still macroscopic and homogeneous) pieces of each individual phase
\cite{guggenheim67}. The most interesting configurations of two coexisting
phases cannot be described by a single canonical ensemble. Important
inter-phase fluctuations remain outside, etc. This is all hidden due to the
restriction to homogeneous systems in the thermodynamic limit.

Also the second law can rigorously be formulated only microcanonically:
Already Clausius \cite{clausius1854} distinguished between external and
internal entropy generating mechanisms. The second law is only related to
the latter mechanism \cite{prigogine71}, the internal entropy generation.
Again, canonical Boltzmann-Gibbs statistics is insensitive to this
important difference.

For this purpose, and also to describe small systems like fragmenting
nuclei or non-extensive ones like macroscopic systems at phase-separation,
or even very large, self-gravitating, systems, we need a new and deeper
definition of statistical mechanics and as the heart of it: of entropy. For
this purpose it is crucial to avoid the thermodynamic limit.

The main aspects of this new thermodynamics were described in
\cite{gross174,gross215,gross214} and especially were introduced to the
chemists community in \cite{gross186,gross213}. I will repeat here only the
basic arguments. Additionally, I will stress the fact that negative heat
capacity and convex entropy can be seen at proper phase transitions of 1.
order, i.e. at phase {\em separation}, in small as well in macroscopic
systems independently whether they have long or short range interactions.
As there was a hot discussion at this conference about this point, it seems
necessary to repeat the arguments here.

\section{What is entropy?} Entropy, S, is the characteristic
entity of thermodynamics. Its use distinguishes thermodynamics from all
other physics; therefore, its proper understanding is essential. The
understanding of entropy is sometimes obscured by frequent use of the
Boltzmann-Gibbs canonical ensemble, and the thermodynamic limit. Also its
relationship to the second law is beset with confusion between external
transfers of entropy $d_eS$ and its internal production $d_iS$.

The main source of the confusion is of course the lack of a clear {\em
microscopic and mechanical} understanding of the fundamental quantities of
thermodynamics like heat, external vs. internal work, temperature, and last
not least entropy, at the times of Clausius and possibly even today.

Clausius \cite{clausius1854} defined  a quantity which he first called the
{\em ``value of metamorphosis'', in German `` Verwandlungswert'' } in
\cite{clausius1854}. Eleven years later he \cite{clausius1865} gave it the
name ``entropy'' $S$:
\begin{equation}
S_b-S_a=\int_a^b{\frac{dE}{T}},\label{entropy}
\end{equation} where $T$ is the absolute temperature of the body when the
momentary change is done, and $dE$ is the increment (positive resp.
negative) of all different forms of energy (heat and potential) put into
resp. taken out of the system.

From the observation that heat does not flow from cold to hot (see section
\ref{zerolaw}, formula \ref{02law}, however section \ref{chsplit}) he went
on to enunciate the second law as:
\begin{equation}
\Delta S=\oint{\frac{dE}{T}}\ge 0,\label{secondlaw}
\end{equation}
which Clausius called the "{\em uncompensated metamorphosis}". As will be
worked out in section \ref{chsplit} the second law as presented by
eq.(\ref{secondlaw}) remains valid even in cases where heat flows from low
to higher temperatures.

Prigogine \cite{prigogine71}, c.f. \cite{guggenheim67}, quite clearly
stated that the variation of $S$ with time is determined by two, crucially
different, mechanisms of its changes: the flow of entropy $d_eS$ to or from
the system under consideration; and its internal production $d_iS$. While
the first type of entropy change $d_eS$ (that effected by exchange of heat
$d_eQ$ with its surroundings) can be positive, negative or zero, the second
type of entropy change $d_iS$ is fundamentally related to its spontaneous
internal evolution (``Verwandlungen'', ``metamorphosis''
\cite{clausius1854}) of the system, and states the universal
irreversibility of spontaneous transitions. It can be only positive in any
spontaneous transformation.

Clausius gives an illuminating example in \cite{clausius1854}: When an
ideal gas suddenly streams under isolating conditions from a small vessel
with volume $V_1$ into a larger one ($V_2>V_1$), neither its internal
energy $U$, nor its temperature changes, nor external work done, but its
internal (Boltzmann-)entropy $S_i$ eq.(\ref{boltzmann0}) rises, by $\Delta
S=N\ln{(V_2/V_1)}$ . Only by compressing the gas (e.g. isentropically) and
creating heat $\Delta E=E_1[(V_2/V_1)^{2/3}-1]$ (which must be finally
drained) it can be brought back into its initial state. Then, however, the
entropy change in the cycle, as expressed by integral (\ref{secondlaw}), is
positive ($=N\ln{(V_2/V_1)}$). This is also a clear example for a
microcanonical situation where the entropy change by an irreversible
metamorphosis of the system is absolutely internal. It occurs  during the
first part of the cycle, the expansion, where there is no heat exchange
with the environment, and consequently no contribution to the
integral(\ref{secondlaw}). The construction by eq.(\ref{secondlaw}) is
correct though artificial. After completing the cycle the Boltzmann-entropy
of the gas is of course the same as initially. All this will become much
more clear by Boltzmann's microscopic definition of entropy, which will
moreover clarify its real {\em statistical} nature:

Boltzmann\cite{boltzmann1872} later defined the entropy of an isolated
system (for which the energy exchange with the environment $d_eQ \equiv 0$)
in terms of the sum of possible configurations, $W$, which the system can
assume consistent with its constraints of given energy  and
volume:\begin{equation} \fbox{\fbox{\vecbm{S=k*lnW}}}\label{boltzmann0}
\end{equation}as written on Boltzmann's tomb-stone, with
\begin{equation}
W(E,N,V)= \int{\frac{d^{3N}\nvec{p}\;d^{3N}\nvec{q}}{N!(2\pi\hbar)^{3N}}
\epsilon_0\;\delta(E-H\{\nvec{q},\nvec{p}\})}\label{boltzmann}
\end{equation} in semi-classical approximation. $E$ is the total energy, $N$ is
the number of particles and $V$ the volume. Or, more appropriate for a
finite quantum-mechanical system:
\begin{equation}W(E,N,V)
=\sum{\scriptsize\begin{array}{ll}\mbox{all eigenstates n of H with given
N,$V$,}\\\mbox{and } E<E_n\le E+\epsilon_0
\end{array}}
\end{equation}
and $\epsilon_0\approx$ the macroscopic energy resolution.  This is still
up to day the deepest, most fundamental, and most simple definition of
entropy. {\em There is no need of the thermodynamic limit, no need of
concavity, extensivity and homogeneity.}  In its semi-classical
approximation, eq.(\ref{boltzmann}), $W(E,N,V,\cdots)$ simply measures the
area of the sub-manifold of points in the $6N$-dimensional phase-space
($\Gamma$-space) with prescribed energy $E$, particle number $N$, volume
$V$, and some other time invariant constraints which are here suppressed
for simplicity. Because it was Planck who coined it in this mathematical
form, I will call it the Boltzmann-Planck principle. It is further
important to notice that $S(E,N,V)$ is everywhere analytical in $E$
\cite{gross206}. In the microcanonical ensemble are no ``jumps'' or
multivaluedness in $S(E)$, independently of whether there are phase
transitions or not, in clear contrast to the canonical $S(T,N,V)$. A fact
which underlines the fundamental role of microcanonical statistics.

The Boltzmann-Planck formula has a simple but deep physical interpretation:
$W$ or $S$  measure our ignorance about the complete set of initial values
for all $6N$ microscopic degrees of freedom which are needed to specify the
$N$-body system unambiguously\cite{kilpatrick67}. To have complete
knowledge of the system we would need to know (within its semiclassical
approximation (\ref{boltzmann})) the initial positions and velocities of
all $N$ particles in the system, which means we would need to know a total
of $6N$ values. Then $W$ would be equal to one and the entropy, $S$, would
be zero. However, we usually only know the value of a few parameters that
change slowly with time, such as the energy, number of particles, volume
and so on. We generally know very little about the positions and velocities
of the particles. The manifold of all these points in the $6N$-dim. phase
space, consistent with the given macroscopic constraints of $E,N,V,\cdots$,
is the microcanonical ensemble, which has a well-defined geometrical size
$W$ and, by equation (\ref{boltzmann0}), a non-vanishing entropy,
$S(E,N,V,\cdots)$. The dependence of $S(E,N,V,\cdots)$ on its arguments
determines completely thermostatics and equilibrium thermodynamics.

Clearly, Hamiltonian (Liouvillean) dynamics of the system cannot create the
missing information about the initial values - i.e. the entropy
$S(E,N,V,\cdots)$ cannot decrease. As has been further worked out in
\cite{gross183,gross207} the inherent finite resolution of the macroscopic
description implies an increase of $W$ or $S$ with time when an external
constraint is relaxed. Such is a statement of the second law of
thermodynamics, which requires that the {\em internal} production of
entropy be positive for every spontaneous process. Analysis of the
consequences of the second law by the microcanonical ensemble is
appropriate because, in an isolated system (which is the one relevant for
the microcanonical ensemble), the changes in total entropy must represent
the {\em internal} production of entropy, see above, and there are no
additional uncontrolled fluctuating energy exchanges with the environment.

\section{The Zero'th Law in conventional extensive Thermodynamics \label{zerolaw}} In
conventional (extensive) thermodynamics thermal equilibrium of two systems
(1 \& 2) is established by bringing them into thermal contact which allows
free energy exchange. Equilibrium is established when the total entropy
\begin{equation}
S_{1+2}(E,E_1)=S_1(E_1)+S_2(E-E_1)\label{eq1}
\end{equation}
is maximal. Under an energy flux $\Delta E_{2\to 1}$ from $2\to 1$ the
total entropy changes to lowest order in $\Delta E$ by
\begin{equation}
\Delta S_{1+2}|_E=(T_2-T_1)\Delta E_{2\to 1}.
\end{equation}
Consequently, a maximum of $S_{total}(E,E_1)|_E\ge S_{1+2}$ will be
approached when
\begin{equation}
 \mbox{sign}(\Delta S_{total})=\mbox{sign}(T_2-T_1)\mbox{sign}(\Delta E_{2\to
 1})>0.\label{02law}
\end{equation}
From here Clausius' first formulation of the Second Law follows: "Heat
always flows from hot to cold". Essential for this conclusion is the {\em
additivity} of $S$ under the split (eq.\ref{eq1}). There are no
correlations, which are destroyed when an extensive system is split.
Temperature is an appropriate control parameter for extensive systems.

It is further easy to see that {\em the heat capacity of an extensive
system with $S(E,N)=2S(E/2,N/2)$ is necessarily positive}
\begin{equation}
C_V(E)=\partial E/\partial T=-~~\frac{(\partial S/\partial
E)^2}{\partial^2S/\partial E^2}>0:\label{posheat}
\end{equation}
The combination two pieces of $N/2$ particles each, but with different
energy per particle, one at $e_a=e_2-\Delta e/2$ and a second at
$e_b=e_2+\Delta e/2$, must lead to $S(E_2,N)\ge S(E_a/2,N/2)+S(E_b/2,N/2)$,
the simple algebraic sum of the individual entropies because by combining
the two pieces one normally looses information. This, however, is equal to
$[S(E_a,N)+S(E_b,N)]/2$, thus $S(E_2,N)\ge[S(E_a,N)+S(E_b,N)]/2$. I.e. {\em
the entropy $S(E,N)$ of an extensive system is necessarily concave,
$\partial^2S/\partial E^2<0$} and eq. \ref{posheat} follows. In the next
section we will see that therefore  {\em extensive systems cannot have
phase separation, the characteristic signal of transition of first order.}
\section{No phase separation without a convex, non-extensive
$S(E)$\label{chsplit}}

At phase separation the weight $e^{S(E)-E/T}$ of the configurations with
energy E in the definition of the canonical partition sum
\begin{equation}
Z(T)=\int_0^\infty{e^{S(E)-E/T}dE}\label{canonicweight}
\end{equation} becomes {\em bimodal}, at the transition temperature it has
two peaks, the liquid and the gas configurations which are separated in
energy by the latent heat. Consequently $S(E)$ must be convex ($\partial^2
S/\partial E^2>0$, like $y=x^2$) and the weight in (\ref{canonicweight})
has a minimum at $E_{min}$ between the two pure phases. Of course, the
minimum can only be seen in the microcanonical ensemble where the energy is
controlled and its fluctuations forbidden. Otherwise, the system would
fluctuate between the two pure phases by an, for macroscopic systems even
macroscopic, energy $\Delta E\sim E_{lat}$ of the order of the latent heat.
The heat capacity is
\begin{equation}
C_V(E_{min})=\partial E/\partial T=-~~\frac{(\partial S/\partial
E)^2}{\partial^2S/\partial E^2}<0.\label{negheatcap}
\end{equation}
I.e. {\em the convexity of $S(E)$ and the negative heat capacity are the
generic and necessary signals of phase-separation\cite{gross174}}. It is
amusing that this fact, which is essential for the original purpose of
Thermodynamics to describe steam engines, seems never been really
recognized in the past 150 years. However,  such macroscopic energy
fluctuations and the resulting negative specific heat are already early
discussed in high-energy physics by Carlitz \cite{carlitz72}.

The existence of the negative heat capacity at phase separation has a
surprising but fundamental consequence: Combining two equal systems with
negative heat capacity and different energy per particle, they will relax
with a flow of energy from the lower to the higher temperature! This is
consistent with the naive picture of an {\em energy equilibration}. Thus
{\em Clausius' "energy flows always from hot to cold", i.e. the dominant
control-role of the temperature in thermo-statistics as emphasized by Hertz
\cite{hertz10a} is violated}. Of course this shows quite clearly that {\em
unlike to extensive thermodynamics the temperature is not the appropriate
control parameter in non-extensive situations like e.g. at phase
separations, nuclear fragmentation, or stellar systems.\cite{gross212}}

By the same reason the well known paradox of Antonov in astro-physics due
to the occurrence of negative heat capacities must be reconsidered: By
using standard arguments from extensive thermodynamics  Lynden-Bell
\cite{lyndenbell68} claims that a system $a$ with negative heat capacity
$C_a<0$ in gravitational contact with another $b$ with positive heat
capacity $C_b>0$ will be unstable: If initially $T_a>T_b$ the hotter system
$a$ transfers energy to the colder $b$ and by this both become even hotter!
If $C_b>-C_a$, $T_a$ rises faster than $T_b$ and if the heat capacities
don't change, this will go for ever. This is Lynden-Bells gravo-thermal
catastrophe. This is wrong because just the opposite happens, the hotter
$a$ may even {\em absorb} energy from the colder $b$ and both systems come
to equilibrium at the same intermediate temperature c.f.
\cite{gross203,gross212}. Negative heat can only occur in the
microcanonical ensemble.

As phase separation exists also in the thermodynamic limit, by the same
arguments as above {\em the curvature of $S(E)$ remains convex,
$\partial^2S/(\partial E)^2>0$. Consequently, the negative heat capacity
should also be seen in ordinary macroscopic systems studied in chemistry!}
see section \ref{macrophase}.

Searching for example in Guggenheims book \cite{guggenheim67} one finds
some cryptic notes in \S 3 that the heat capacity of steam at saturation is
negative. No notice that {\em this is the generic effect at any phase
separation!} Therefore let me recapitulate in the next section how chemists
treat phase separation of macroscopic systems and then point out why this
does not work in non-extensive systems like fragmenting nuclei, at phase
separation in normal macroscopic systems, or large astronomical systems.
\section{Macroscopic systems in Chemistry \label{chemistry}}

Systems studied in chemical thermodynamics consist of several {\em
homogeneous macroscopic} phases $\alpha_1,\alpha_2,\cdots$
cf.\cite{guggenheim67}. Their mutual equilibrium must be explicitly
constructed from outside.

Each of these phases are assumed to be homogeneous and macroscopic (in the
"thermodynamic limit" ($N_\alpha\to\infty|_{\rho_\alpha=const}$)). There is
no common canonical ensemble for the entire system of the coexisting
phases. Only the canonical ensemble of {\em each} phase separately becomes
equivalent in the limit to its microcanonical counterpart.

The canonical partition sum of {\em each} phase $\alpha$ is defined as the
Laplace transform of the underlying  microcanonical sum of states
$W(E)_\alpha=e^{S_\alpha(E)}$ \cite{gross147,gross158}
\begin{equation}
Z_\alpha(T_\alpha)= \int_0^\infty e^{S_\alpha(E)-E/T_\alpha} dE.
\end{equation}
The mean canonical energy is
\begin{equation}
 <E_\alpha(T_\alpha)>=T_\alpha^2\partial \ln Z_\alpha(T_\alpha)/\partial T_\alpha.
\end{equation}
In chemical situations proper the assumption of homogeneous macroscopic
individual phases is of course acceptable. In the thermodynamic limit
($N_\alpha\to\infty|_{\rho_\alpha=const}$) of a {\em homogeneous} phase
$\alpha$, the canonical energy\\ $<\!\!E_\alpha(T_\alpha)\!\!>$ becomes
identical to the microcanonical energy $E_\alpha$ when the temperature is
determined by $ T_\alpha^{-1}=\partial S_\alpha(E,V_\alpha)/\partial
E_\alpha$. The relative width of the canonical energy is
\begin{equation}
\Delta
E(T)_\alpha=\frac{\sqrt{<E_\alpha^2>_T-<E_\alpha>_T^2}}{<E_\alpha>_T}\propto
\frac{1}{\sqrt{N_\alpha}}.
\end{equation}
The heat capacity at constant volume is
\begin{eqnarray}
C_\alpha|_{V_\alpha}&=&\frac{<E_\alpha^2>_{T_\alpha}-<E_\alpha>_{T_\alpha}^2}{T_\alpha^2}\ge
0.\label{specheat}
\end{eqnarray}

Only in the thermodynamic limit ($N_\alpha\to\infty|_{\rho_\alpha=const}$)
does the relative energy uncertainty $\Delta E_\alpha\rightarrow 0$, and
the canonical and the microcanonical ensembles for each homogeneous phase
($\alpha$) become equivalent. This equivalence is the only justification of
the canonical ensemble controlled by intensive temperature $T$, or chemical
potential $\mu$, or pressure $P$. I do not know of any microscopic
foundation of the canonical ensemble and intensive control parameters apart
from the limit.

The positiveness of any canonical $C_V(T)$ or $C_P(T)$ (\ref{specheat}) is
of course the reason why the inhomogeneous system of several coexisting
phases ($\alpha_1 \& \alpha_2$) with an overall {\em negative} heat
capacity cannot be described by a {\em single common} canonical
distribution \cite{gross159,gross174}.

\section{New kind of phases well seen in hot nuclei
or multi-star systems.\label{nuclearfrag}}

The new lesson to be learned is that if one defines the phases by
individual peaks \footnote{Here I do not mean irregularities of the order
of $N^{-1/3}$ due to the discreteness of the quantum level distributions}
in $e^{S(E)-E/T}$ in (\ref{canonicweight}), then there exist also {\em
inhomogeneous phases} like in fragmented nuclei or stellar systems. The
general concept of thermo-statistics becomes enormously widened.

Now, certainly neither the phase of the whole multi-fragmented nucleus nor
the individual fragments themselves can be considered as macroscopic
homogeneous phases in the sense of chemical thermodynamics (ChTh).
Consequently, (ChTh) cannot and should not be applied to fragmenting nuclei
and the microcanonical description is ultimately demanded. This becomes
explicitly clear by the fact that the configurations of a multi-fragmented
nucleus have a {\em negative} heat capacity at constant volume $C_V$ and
also at constant pressure $C_P$ (if at all a pressure can be associated to
nuclear fragmentation \cite{gross174}). Meanwhile experimental evidences of
negative heat capacities have accumulated: Nuclear fragmentation e.g.
\cite{dAgostino00}, atomic clusters e.g. \cite{schmidt01}, astrophysics
e.g. \cite{thirring70}, conventional macroscopic systems at phase
separation e.g.\cite{guggenheim67}.

The existence of well defined peaks (i.e. phases as defined above) in the
event distribution of nuclear fragmentation data is demonstrated very
nicely in \cite{pichon03} from various points of view. A lot more physics
about the mechanism of phase transitions can be learned from such studies.
\section{Application in astrophysics}
The necessity of using ``extensive'' instead of ``intensive'' control
parameter is explicit in astrophysical problems. E.g.: for the description
of rotating stars one conventionally works at a given temperature  and
fixed angular velocity $\Omega$ c.f. \cite{chavanis03}. Of course in
reality there is neither a heat bath nor a rotating disk. Moreover, the
latter scenario is fundamentally wrong as at the periphery of the disk the
rotational velocity may even become larger than velocity of light.
Non-extensive systems like astro-physical ones do not allow a
``field-theoretical'' description controlled by intensive fields !

E.g. configurations with a maximum of random energy
\begin{equation}
E_{random}=E-\frac{\Theta\Omega^2}{2} -E_{pot}
\end{equation} and consequently with the largest entropy are the ones
with smallest moment of inertia $\Theta$ compact single stars. Just the
opposite happens when the angular-momentum $L$ and not the angular velocity
$\Omega$ are fixed:\begin{equation} E_{random}=E-\frac{L^2}{2 \Theta}
-E_{pot}.
\end{equation}Then configurations with large moment of inertia are
maximizing the phase space and the entropy. I.e. eventually double or multi
stars are produced, as observed in reality.

\begin{figure}[h]
\hspace*{-0.5cm}
\includegraphics[bb =0 0 511 353,width=8cm,angle=0,clip=true]{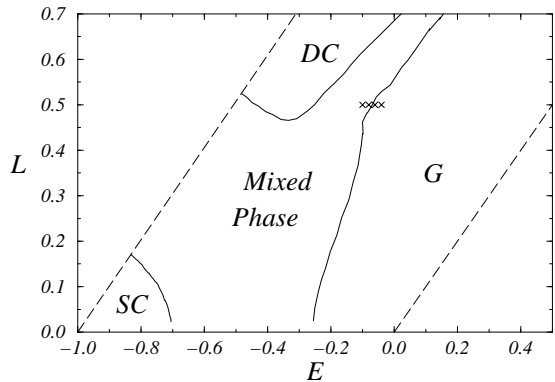}
\caption{Phase diagram of rotating self-gravitating systems in the
energy-angular-momentum $(E,L)$-plane. DC: region of double-stars, G: gas
phase, SC: single stars. In the mixed region one finds various exotic
configurations like ring-systems in coexistence with gas, double stars or
single stars. In this region of phase-separation the heat capacity is
negative and the entropy is convex. The dashed lines $E-L=1$ (left) and
$E=L$ (right) delimit the region where calculations were carried
out.\label{phased}}
\end{figure}

In figure \ref{phased} one clearly sees the rich and realistic
microcanonical phase-diagram of a rotating gravitating system controlled by
the ``extensive'' parameters energy and angular-momentum. \cite{gross187}

\section{Negative heat capacity at phase-separation can also be seen in macroscopic systems
independently of the range of the interaction.\label{macrophase} }

The``convex intruder'' in $S(E)$ with the depth $\Delta S_{surf}(E_{min})$
has a direct physical significance: Its depth is the surface entropy due to
constraints by the existence of the inter-phase boundary between the
droplets of the condensed phase and the gas phase and the corresponding
correlation. $\Delta S_{surf}(E_{min})$ is directly related to the surface
tension per surface atom (with number $N_{surf}$) of the droplets.
\begin{equation}
\sigma_{surf}/T_{tr}=\frac{\Delta S_{surf}(E_{min})}{N_{surf}}
\end{equation}

In my paper together with M.Madjet \cite{gross157} we have compared the
values of $\Delta S_{surf}(E_{min})$ calculated by Monte-Carlo using a
realistic short range interaction with the values of the surface tension of
the corresponding macroscopic system.
\begin{figure}[h]\cent{
\includegraphics*[bb = 99 57 400 286, angle=-0, width=8cm,
clip=true]{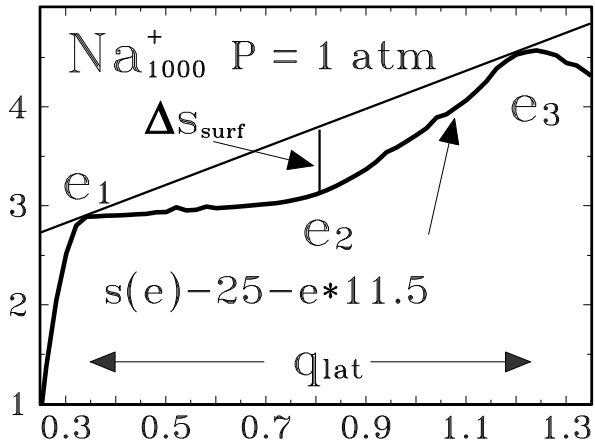}} \caption{Microcanonical Monte-Carlo
(MMMC)~\protect\cite{gross157,gross174} simulation of the entropy
  \index{entropy} $s(e)$ per atom ($e$ in eV per atom) of a system of
  $N=1000$ sodium atoms at an external pressure of 1 atm.  At the
  energy $e\leq e_1$ the system is in the pure liquid phase and at
  $e\geq e_3$ in the pure gas phase, of course with fluctuations. The
  latent heat per atom is $q_{lat}=e_3-e_1$.  \underline{Attention:}
  the curve $s(e)$ is artificially sheared by subtracting a linear
  function $25+e*11.5$ in order to make the convex intruder visible.
  {\em $s(e)$ is always a steep monotonic rising function}.  We
  clearly see the global concave (downwards bending) nature of $s(e)$
  and its convex intruder. Its depth is the entropy loss due to
  additional correlations by the interfaces. It scales $\propto
  N^{-1/3}$. From this one can calculate the surface
  tension per surface atom
  $\sigma_{surf}/T_{tr}=\Delta s_{surf}*N/N_{surf}$.  The double
  tangent (Gibbs construction) is the concave hull of $s(e)$. Its
  derivative gives the Maxwell line in the caloric curve $e(T)$ at
  $T_{tr}$. In the thermodynamic limit the intruder would disappear
  and $s(e)$ would approach the double tangent from below, not of course
  $S(E)$, which remains deeply convex: The probability of configurations with
  phase-separations is suppressed by the (infinitesimal small)
  factor $e^{-N^{2/3}}$ relative to the pure phases and the
  distribution remains {\em strictly bimodal in the canonical ensemble}
  in which the region $e_1<e<e_3$ of phase separation gets lost.\label{naprl0f}}
\end{figure}
Table (\ref{table}) shows the scaling behavior of $\Delta
S_{surf}(E_{min})$ with the size $N$ of the system.
\begin{table}[h]
\caption{Parameters of the liquid--gas transition of small
  sodium clusters (in the MMMC-calculation~\protect\cite{gross157,gross174}
  several clusters coexist) in
  comparison with the bulk for a rising number $N$ of atoms,
  $N_{surf}$ is the average number of surface atoms (estimated here as
  $\sum{N_{cluster}^{2/3}}$) of all clusters with $N_i\geq2$ together.
  $\sigma/T_{tr}=\Delta s_{surf}*N/N_{surf}$ corresponds to the
  surface tension. Its bulk value is adjusted to agree with the
  experimental values of the $a_s$ parameter which
  we used in the liquid-drop formula for the binding energies of small
  clusters, c.f.  Brechignac et al.~\protect\cite{brechignac95}, and
  which are used in this calculation~\cite{gross174} for the individual
  clusters.\label{table}}
\begin{center}
\renewcommand{\arraystretch}{1.4}
\setlength\tabcolsep{5pt}
\begin{tabular} {|c|c|c|c|c|c|} \hline
&$N$&$200$&$1000$&$3000$&\vecb{bulk}\\ 
\hline \hline &$T_{tr} \;[K]$&$940$&$990$&$1095$&\vecb{1156}\\ \cline{2-6}
&$q_{lat} \;[eV]$&$0.82$&$0.91$&$0.94$&\vecb{0.923}\\ \cline{2-6} {\bf
Na}&$s_{boil}$&$10.1$&$10.7$&$9.9$&\vecb{9.267}\\ \cline{2-6} &$\Delta
s_{surf}$&$0.55$&$0.56$&$0.44$&\\ \cline{2-6}
&$N_{surf}$&$39.94$&$98.53$&$186.6$&\vecbm{$\infty$}\\ \cline{2-6}
&$\sigma/T_{tr}$&$2.75$&$5.68$&$7.07$&\vecb{7.41}\\ \hline
\end{tabular}
\end{center}
\end{table}

Roughly $\Delta S_{surf}(E)\propto N^{2/3}$ and one may argue that this
will vanish compared to the ordinary leading volume term $S_{vol}(E)\propto
N$. However, this is not so as $S_{vol}(E)$ at energies inside the
phase-separation region (the convex intruder) is the \underline{concave
hull} of $S(E)$ (its slope gives the Maxwell construction of the caloric
curve $T(E)$). It is a straight line and its curvature $\partial^2
S_{vol}/\partial E^2\equiv 0$. Consequently for large $N$
\begin{eqnarray}
\partial^2S/\partial E^2&\sim&
\partial^2 S_{vol}/\partial E^2+\partial^2\Delta S_{surf}/\partial
E^2 +\cdots\nonumber\\ &\asymp & \partial^2\Delta S_{surf}/\partial E^2
\end{eqnarray} and the depth of the intruder $\Delta S_{surf}(E_{min})=\sigma/T_{tr}
*N_{surf}\sim N^{2/3}$ goes to infinity in the thermodynamic limit. Of
course, the ubiquitous phenomena of phase separation exist only by this
reason. It determines the (negative) heat capacity as in
eq.(\ref{negheatcap}). The physical (quite surprising) consequences are
discussed in \cite{gross214,gross213}.

\section{The microscopic origin of the convexities of {\boldmath$S(E)$} and of
phase-separation\label{convex}} Many
applications of microcanonical thermodynamics to realistic examples of hot
nuclei, atomic clusters, and rotating astrophysical systems have been
presented during the past twenty years which demonstrate convex intruders
in the microcanonical entropy and, consequently, negative heat capacities.
Such are reviewed in the publication list on the web site
http://www.hmi.de/people/gross/ and
elsewhere\cite{chomaz99,chomaz00,chomaz00a}. Here we shall illuminate the
general microscopic mechanism leading to the appearance of a convex
intruder in $S(E,V,N,,\cdots)$ as far as possible by rigorous and
analytical methods. This is the generic signal of phase transitions of
first order and of phase-separation within the microcanonical ensemble.
Assume the system is classical and obeys the Hamiltonian:
\begin{eqnarray}H &= &\sum_i^N{\frac{p_i^2}{2m}}+\Phi^{int}[\{\nvec{r}\}]
\label{hamiltonian}\\\nonumber\\\nonumber\\
\Phi^{int}[\{\nvec{r}\}]&:=&\sum_{i<j}{\phi(\nvec{r}_i-\nvec{r}_j)}\nonumber
\end{eqnarray}
In this case the system is controlled by energy and volume. For what
follows it is important to stress that $\Phi^{int}[\{\nvec{r}\}]$ is {\em
independent} of the energy as control parameter. The topology of its
attractive pockets is also independent of the available volume $V$.
\subsection{Liquid-gas transition} The microcanonical sum of states or partition sum is:
\begin{eqnarray}
W(E,N,V)&=&\frac{1}{N!(2\pi\hbar)^{3N}}\times\label{W}\\
\lefteqn{\hspace{-3cm}\int_{V^N}{d^{3N}\nvec{r} \int{d^{3N}\nvec{p}_i{
\epsilon_0\;\delta(E-\sum_i^N{\frac{\nvec{p}_i^2}{2m_i}-\Phi^{int}[\{\nvec{r}\}]})}}}}
\nonumber\\\nonumber
\\&=&\nonumber\\\nonumber\\\nonumber\\\lefteqn{\hspace{-2cm}\frac{V^N
\epsilon_0(E-E_0)^{(3N-2)/2} \prod_1^N{m_i^{3/2}}}{N!\Gamma(3N/2)
(2\pi\hbar^2)^{3N/2}}\;\;\times\nonumber}\\\nonumber\\
\lefteqn{\hspace{-2cm}\int_{V^N}{\frac{d^{3N}r}
{V^N}}\left(\frac{E-\Phi^{int}[\{\nvec{r}\}]}{E-E_0}\right)^{(3N-2)/2}
\label{split}}\\\nonumber\\\nonumber\\
&=&\nonumber\\\nonumber\\\lefteqn{\hspace{-3cm}W_{id-gas}(E-E_0,N,V)\times
W_{conf}(E-E_0,N,V)}\nonumber\\\nonumber\\\nonumber\\&=&
e^{[S_{id-gas}+S_{conf}]}\label{micromeg}\\\nonumber
\end{eqnarray}\
\begin{eqnarray}
W_{id-gas}(E,N,V)&=&\frac{V^N\epsilon_0
E^{(3N-2)/2}\prod_1^N{m_i^{3/2}}}{N!\Gamma(3N/2)
(2\pi\hbar^2)^{3N/2}}\nonumber\\\label{idgas}\\\nonumber\\\nonumber\\
W_{conf}(E-E_0,N,V)&=&\int_{V^N}{\frac{d^{3N}r}
{V^N}}\Theta(E-\Phi^{int}[\{\nvec{r}\}])\nonumber\\ \lefteqn{\hspace{-1cm}
\times\left(1-\frac{\Phi^{int}[\{\nvec{r}\}]-E_0}{E-E_0}\right)^{(3N-2)/2}
\label{Win1}}
\end{eqnarray}
V is the spatial volume; $E_0=\min \Phi^{int}[\{\nvec{r}\}]$ is the energy
of the ground-state of the system. The separation of $W(E,N,V)$ into
$W_{id-gas}$ and $W_{conf}$ is the microcanonical analogue of the split of
the canonical partition sum into a kinetic part and a configuration part:
\begin{equation}
Z(T)=\frac{V^N}{N!}\left(\frac{m
T}{2\pi\hbar^2}\right)^{3N/2}\int{\frac{d^{3N}r}{V^N}e^{-\frac{\Phi^{int}[\{\nvec{r}\}]}{T}}}\label{canonical}
\end{equation}

In the thermodynamic limit, the order parameter of the (homogeneous)
liquid-gas transition is the density. The transition is linked to a
condensation of the system towards a larger density controlled by pressure.
For a finite system, we expect analogous behavior. However, for a closed
finite system, which is allowed to become inhomogeneous at phase
separation, this is controlled by the available system volume $V$ and not
by intensive density or pressure. At low energies, the $N$ particles
condensate into a droplet with much smaller volume $V_{0,N}\ll V$. $3(N-1)$
internal coordinates are limited to $V_{0,N}$. Only the center of mass of
the droplet can move freely in $V$ (remember we did not fix the
center-of-mass in equation eq.(\ref{W})). The system does not fill the
$3N$-configuration space $V_N$. Only a stripe with width $ V_{0N}^{1/3}$ in
$3(N-1)$ dimensions of the total $3N$-dim space is populated. The system is
non-homogeneous even though it is equilibrized and, at low energies,
internally in the single liquid phase; and it is not characterized by an
intensive homogeneous density. In fact, $W_{conf}(E-E_0,N,V)$ can be
written as:
\begin{eqnarray}
W_{conf}(E-E_0,N,V)&=& \left[\frac{V(E,N)}V\right]^N\le 1 \label{Win1b}\\
\left[V(E,N)\right]^N&\stdef&\nonumber\\
 \lefteqn{\hspace{-3cm}\int_{V^N}d^{3N}r\;\Theta(E-\Phi^{int}[\{\nvec{r}\}])}\nonumber\\
\lefteqn{\hspace{-3cm}\times\left(1-\frac{\Phi^{int}[\{\nvec{r}\}]-E_0}{E-E_0}\right)^{(3N-2)/2}}
\label{Sin2}\\\nonumber\\
S_{conf}(E-E_0,N,V)&=&N\ln\left[\frac{V(E,N)}{V}\right]\le0\label{Sin1}
\end{eqnarray}
The first factor $\Theta(E-\Phi^{int}[\{\nvec{r}\}])$ in eq(\ref{Sin2})
eliminates the energetically forbidden regions. Only the potential holes
(clusters) in the $3N$-dim potential surface $\Phi^{int}[\{r\}]\le E$
remain. Their volume $V^N(E,N)\le V^N$ is the accessible part of the
$3N$-dim-spatial volume where $\Phi^{int}[\{r\}]\le E$. I.e. $ V^N(E,N)$ is
the total $3N$-dim. eigen-volume of the condensate (droplets), with $N$
particles at the given energy, summed over all possible partitions,
clusterings, in $3N$-configuration space. The relative volume fraction of
each partition compared with $V^N(E,N)$ gives its relative probability. $
V^N(E,N)$ has the limiting values:
\begin{equation}
{[V(E,N)]^N}=\left\{\begin{array}{ll}V^N&\mbox{for $E$ in the gas
phase}\nonumber\\{V_{0N}}^{N-1}V&\mbox{for~~}E=E_0
\end{array}\right.\label{volume}\end{equation}
$W_{conf}(E-E_0,N,V)$ and $S_{conf}(E-E_0,N,V)$ have the limiting values:
\begin{eqnarray} W_{conf}(E-E_0)&\le& 1,\;\Rightarrow S_{conf}(E-E_0,N)
 \le0\nonumber\\
&\rightarrow& \left\{\begin{array}{ll}1 &\;\;\; \;\;\;\;\;\; \;\;\;E\gg
\Phi^{int}\nonumber\\\left[\frac{V_{0N}}V\right]^{(N-1)}&\;\;\;
\;\;\;\;\;\; \;\;\;E\to E_0\end{array}\right.
\\\\\nonumber\\ S_{conf}(E-E_0)&\to
&\left\{\begin{array}{ll}0&E\gg
\Phi^{int}\nonumber\\ln\left\{[\frac{V_{0N}}V]^{N-1}\right\}< 0&E\to
E_0\end{array}\right. \label{Sin2b}\\\end{eqnarray}

All physical details are contained in $W_{conf}(E-E_0,N,V)$ alias
$N\ln[V(E,N)]$, c.f. eqs.(\ref{Win1b}--\ref{Sin2b}): If the energy is high
the detailed structure of $\Phi^{int}[\{\nvec{r}\}]$ is unimportant
$W_{conf}\approx 1$, $S_{conf}\approx 0$. The system behaves like an ideal
gas and fills the volume $V$. At sufficiently low energies only the minimum
of $\Phi^{int}[\{\nvec{r}\}]$ is explored by $W_{conf}(E-E_0,N, V)$. The
system is in a condensed phase, a single liquid drop, which moves freely
inside the empty larger volume $V$, the $3(N-1)$ internal degrees of
freedom are trapped inside the {\em reduced} volume $V_{0N} \ll V$.

One can guess the general form of $N\ln[V(E,N)]$: Near the groundstate
$E\goo E_0$ it must be flat $\approx(N-1)\ln[V_{0N}]+\ln[V-V_{0N}]$ because
the liquid drop has some eigen-volume $V_{0N}$  in which each particle can
move (liquid). With rising energy $\ln[ V(E,N)]$ rises up to the point
($E_{trans}$) where it is possible that the drop fissions into two. Here an
additional new configuration opens in $3N$-dim configuration space: Either
one particle evaporates from the cluster and explores the external volume $
V$, or the droplet fissions into two droplets and the two CM coordinates
explore the larger $V$. This gives a sudden jump in $S_{conf}(E)$ by
something like $\sim \ln\{\frac{V-V_{0(N-1)}}{V_{0(N-1)}}\}$ and similar
jump upwards in the second case.

Later further such "jumps" may follow. Each of these "jumps" induce a
convex upwards bending of the total entropy $S(E)$ (eq.\ref{micromeg}).
Each is connected to a bifurcation and bimodality of $e^{S(E)-E/T}$ and the
phenomenon of {\em phase-separation}.

In the conventional canonical picture for a large number of particles this
is forbidden and hidden behind the familiar Yang-Lee singularity of the
liquid to gas phase transition. In the microcanonical ensemble this is
analogue to the phenomenon of multi-fragmentation in nuclear systems
\cite{gross174,gross153}. This, in contrast to the mathematical Yang-Lee
theorem, physical microscopic explanation of the liquid to gas phase
transition sheds sharp light on the physical origin of the transition, the
sudden change in the inhomogeneous population of the $3N$-dim.
configuration space.

\subsection{Solid-liquid transition} In contrast to the liquid
phase, in the crystal phase a molecule can only move locally within its
lattice cage of the size $d^3$, instead freely in the whole volume $V_{0N}$
of the liquid condensate. These are deep localized holes in the potential
surfaces of $\Phi^{int}[\{\nvec{r}\}]$. I.e. in equation (\ref{Sin2b})
instead we have $S_{conf}\to\ln\{[\frac{d^3}{V_{0N}}]^{N-1}\}$. The
convexity of $S_{conf}(E)$ is not controlled by the available volume $V$ of
the system.

\subsection{Conclusion}

Only by treating the many-body system microcanonically the $\sim 170$ years
old challenge of thermodynamics is microscopically solved:

The essential differences between the gas, the liquid, and solid phase are
the following: Whereas the gas occupies the whole container, the liquid is
confined to a definite condensate volume, however this may have any shape.
It is separated from the gas by a surface. The solid is also confined to
definite volume but in contrast to the liquid its surface has also a
definite shape. Of course {\em this is the standard experimental
identification of the phase transition and not any singularity (Yang-Lee)}.
However, these differences cannot be seen in the canonical ensemble. As
demonstrated in section \ref{macrophase}, microcanonically this can well be
seen even in macroscopic systems with any kind of interaction, short- and
long range.

The gas- liquid transition is linked to the transition from uniform filling
of the container volume $V$ by the gas to the smaller eigen-volume of the
system $V_0$ in its condensed phase where the system is inhomogeneous (some
liquid drops inside the larger empty volume $V$). First $3(N-1)$, later at
higher energies less and less degrees of freedom condensate into the drop.
First three, then more and more degrees of freedom
(center-of-mass-coordinates of the drops) explore the larger container
volume $V$ leading to upwards jumps (convexities) of $S_{conf}(E$). The
volume of the container controls how close one is to the critical end-point
of the transition, where phase-separation disappears. Towards the critical
end-point, i.e. with smaller V, the jumps $\ln[V-V_0]-\ln[V_0]$ become
smaller and smaller. At the surface of a drop $ \Phi^{int}> E_0=\min
\Phi^{int}$, i.e. the surface gives a negative contribution to $S_{conf}$
in equation (\ref{Sin2}) and to $S$ at energies $E\goo E_0$.

In the case of the solid-liquid transition, however, the external volume,
$V$, of the container confines only the center-of-mass position of the
crystal, resp., the droplet. The entropy jumps during melting by $\Delta
S_{conf}\propto\ln[V_{0N}]-\ln{d^3}$. These jumps are not controlled by the
external volume $V$ resp. the external pressure. There is no critical
end-point of the solid-liquid transition-line in the $V-E$ or $P-E$ phase
diagram.

Discussions with S.Ruffo are gratefully acknowledged.


\end{document}